\documentclass{optica-article}

\journal{opticajournal} 

\articletype{Research Article}

\usepackage{lineno}
\usepackage[normalem]{ulem}
\usepackage{slashbox}

\begin{document}

\title{Blue-shifted dispersive waves and broadband UV emission using dual-core SiN waveguides} 

\author{L. Xia,\authormark{1,*} P.J.M. van der Slot,\authormark{1,2} M. Timmerkamp,\authormark{3} C. Fallnich,\authormark{3} and K.-J. Boller\authormark{1,3}}

\address{\authormark{1}Laser Physics and Nonlinear Optics, Department of Applied Nanophotonics, Faculty of Science and Technology, MESA+ Institute, University of Twente, P.O. Box 217, 7500 AE Enschede, the Netherlands\\}
\address{\authormark{2}Nonlinear Nanophotonics, Department of Applied Nanophotonics, Faculty of Science and Technology, MESA+ Institute, University of Twente, P.O. Box 217, 7500 AE Enschede, the Netherlands\\}
\address{\authormark{3}Optical Technologies Group, Institute of Applied Physics, University of Münster, Corrensstraße 2, 48149 Münster, Germany\\}

\email{\authormark{*}l.xia@utwente.nl} 


\begin{abstract*}
We show that using strongly coupled dual-core waveguides for supercontinuum generation shifts the wavelength of the high-frequency dispersive waves towards shorter wavelengths, as compared to generation in a single-core waveguide having the same core dimensions. In a demonstration experiment, we launch ultrashort infrared pump pulses at 1-$\mu$m wavelength (285-THz frequency) into silicon nitride waveguides, where soliton formation and fission leads to generation of dispersive waves in the visible range. Efficient input coupling and controlled excitation of the fundamental supermodes of the dual-core waveguide is provided with adiabatic tapers and a dual-prong input structure. For the dual-core waveguide, the short-wavelength dispersive wave is located at 540~nm (green, 555~THz), which is blue-shifted by 80~nm (70~THz) compared to that of the single-core waveguide. Simultaneously, the dual-core waveguide generates broadband radiation spanning from the blue into the UV range, reaching to below 350~nm (above 855~THz), with typically a spectral density 25~dB below that of the dispersive wave. The broadband component can be addressed to third harmonic generation and is not observed in single-core supercontinuum generation. Numerical modeling shows good agreement with experimental measurements. The demonstrated dual-core approach and dedicated input coupling appear to hold promise also for other waveguide structures, independent of specific materials or core dimensions, by providing shorter wavelengths than with the respective single-core waveguide.
\end{abstract*}

\section{Introduction}
Controlling the spectral shape of supercontinuum (SC) generation, also named white-light generation, is of increasing importance in modern optics and photonics. Initially, tapered and photonic crystal fibers opened up more options for dispersion engineering, a prerequisite to increase spectral bandwidth and shape output spectra~\cite{dudley2006supercontinuum,petersen2017increased,yu2025visible}. With the advent of low-loss and high index-contrast waveguides~\cite{blumenthal2018silicon,hu2007low,robinson2008first}, SC generation entered a novel phase~\cite{bres2023supercontinuum,dutt2024nonlinear,xu2022towards, sirleto2023introduction}. The much stronger optical confinement reduced power requirements and enabled coherent generation via much faster nonlinear dynamics and, correspondingly, shorter interaction lengths~\cite{Torres-CompanyDOI10.1364/OE.450987}. In addition to these advantages, integrated waveguides offer an additional degree of freedom in shaping and broadening the output spectra via the transverse cross-sectional design of waveguides. For example, providing broadband and strong anomalous dispersion in thick and wide-core silicon nitride (SiN) waveguides enabled SC generation with extreme optical bandwidth~\cite{Epping2015,PorcelDOI10.1364/OE.25.001542,liu2016octave}. Furthermore, the cross section can be designed to vary along the propagation direction to create optical components such as waveguide tapers~\cite{mu2020edge}, beam splitters~\cite{amanti2024integrated}, or resonators~\cite{zhang2024advances}. 

Of particular focus in cross-sectional design is controlling the generation of dispersive waves, which are emitted by ultrashort soliton pulses~\cite{castello2014inverse,turner2006tailored,mas2010tailoring} that can form if the waveguide is engineered to impose anomalous group velocity dispersion (GVD). Interest in dispersive waves is high because the corresponding wavelengths lie at the very outer wings of the SC spectra where they provide high spectral intensities. To exploit these properties, initially, dispersive wave generation targeted mostly the infrared range, for applications such as chip-integrated molecular detection~\cite{guo2020nanophotonic,lee2024chip, DiXia2021chip} or to address the telecom wavelength range for signal processing~\cite{Ettabib15}.

More recently, interest has increased in controlling dispersive wave generation towards shorter wavelengths to serve applications such as involving biomedical imaging~\cite{thimsen2017shortwave,cui2017visible}, optical atomic clocks~\cite{lee2017towards,moille2025broadband,li2025generation} or quantum photonics~\cite{zhao2020visible,ji2025dispersive}. 

However, shifting dispersive waves toward visible or even UV wavelengths turned out to be much more challenging. The reason is that in these ranges the refractive index of the involved core and cladding materials rises steeply which affects phase matching, i.e., the required matching of the phase velocity of dispersive waves to the group velocity of the radiating solitons. In spite of this problem, in fiber systems, tuning dispersive waves to shorter wavelengths was demonstrated, caused by the Raman-induced red-shift of the solitons which increases with fiber lengths~\cite{yuan2014blue,tartara2003blue,cristiani2004dispersive}, or by increasing the Kerr contribution in the phase matching condition for the dispersive waves ~\cite{yan2016visible}.

In integrated waveguides, a range of approaches for shifting dispersive waves towards shorter wavelengths have been studied as well, experimentally and numerically. For instance, experimental demonstration confirmed that dispersive waves can be generated at shorter wavelengths via a cascaded process with pumping in the normal group velocity dispersion regime~\cite{okawachi2017coherent}. However, because such dynamics occurs in two stages, the process is typically less efficient than when pumped directly in the anomalous GVD regime, such as one order of magnitude when compared to~\cite{Tagkoudi21}. Numerical modeling uncovered another way to control dispersive wave generation using two coupled resonators, where the coupler length is varied to tune the spectral splitting of the resonator modes and their dispersion~\cite{dorche2017extending}. In several experiments, dispersive waves were created at shorter wavelengths by increasing the index contrast via using air as top and side cladding~\cite{Tagkoudi21,Lafforgue20,Lu19,yoon2017coherent}. However, this approach is less general as it renders the un-clad waveguides more susceptible to environmental influence. Another approach was to manipulate dispersive waves through quasi-phase matching via spatially modulating the dispersion along propagation~\cite{yang2024generation}. However, the approach needs relatively long interaction lengths because periodic adiabatic tapering is required between sections of different waveguide cross sections. Recently, based on numerical investigations, it was shown that slot waveguides can be used for shifting dispersive waves into the blue range~\cite{das2021dispersion,wang2021efficient,fang2022multiple}. However, the suggested design uses horizontally oriented slots, therefore excitation of the desired spatial mode remains difficult. The latter may also explain  why so far no visible dispersive waves were observed in slot waveguides.

What would be required is an approach that blue-shifts the shortest zero-dispersion wavelength (ZDW) of a given waveguide. Ideally, the approach would also have to support waveguided input coupling to the desired mode.  

Here we demonstrate blue shifting of the short-wavelength dispersive wave in integrated waveguides via an approach that is based on using strong coupling to an additional waveguide. The strong coupling can be viewed to split the single-core fundamental mode into two new supermodes that are spatially symmetric (SM) and anti-symmetric (AM)~\cite{chuang1987application,hardy2003coupled,huang1994coupled,de2010supermode}, and which have strongly different dispersion. One of the modes, which is the anti-symmetric supermode,  exhibits a higher-frequency ZDW, and increases the strength and spectral width of anomalous GVD. Specifically, we generate supercontinuum in the anti-symmetric supermode of a dual-core silicon nitride waveguide structure. Efficient and integrated excitation of the supermode is achieved with a longitudinal circuit design that incorporates a fork-shaped, dual-prong input waveguide section equipped with inverse tapers with a measured coupling efficiency of 80~$\%$. Injecting ultrashort pump pulses with a 100-fs width at 1-$\mu$m wavelength into the dual-core structure, we generate dispersive waves in the green spectral range, at 540~nm. This corresponds to a shift of the dispersive wave by 80~nm towards shorter wavelengths as compared to a single-core reference waveguide with the same cross section. The dual-core structure also generates broadband blue and UV radiation, typically 25~dB below the peak of the dispersive wave, with the central wavelength shifting from 410~nm into the UV below 350~nm when increasing the pump power.

\section{Methods}
\label{Methods}
\subsection{Waveguides \label{waveguides}}

\begin{figure}[b!]
\centering
\includegraphics[width=1\linewidth]{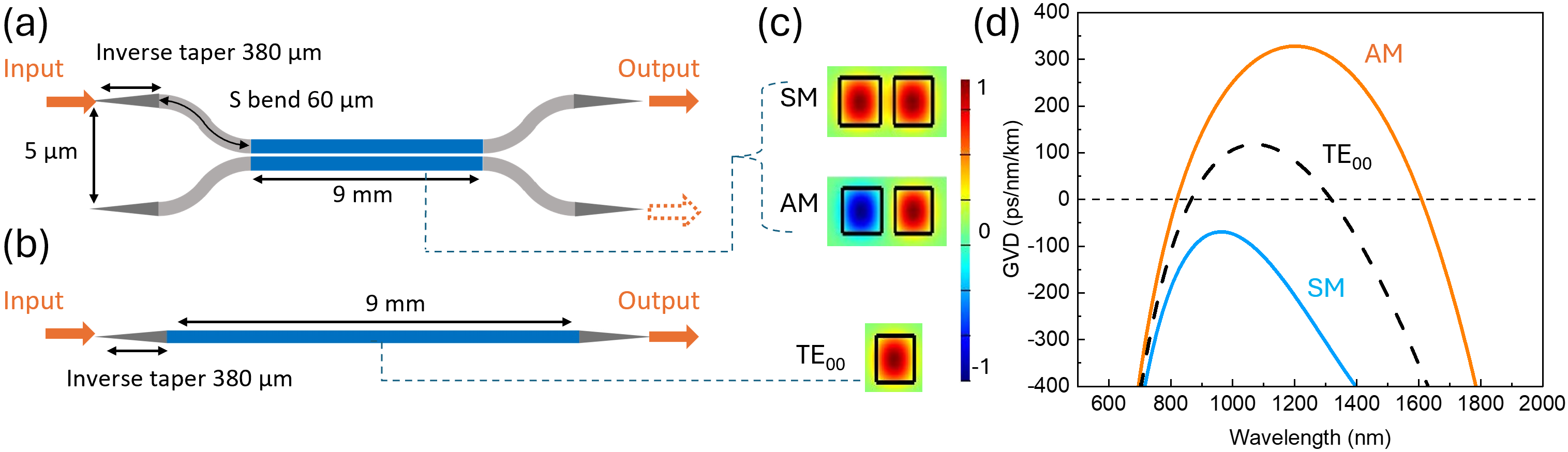}
\caption{(a) Schematic top view  of the dual-core sample. The prongs are equipped with 380~$\mu$m-long inverse tapers providing a MFD of 2~$\mu$m at the facets. The part colored in blue is the strongly coupled dual-core waveguide section with a small gap of 300~nm between the two cores, each 0.8-$\mu$m wide and 0.8-$\mu$m thick. (b) Single-core reference sample having the same core cross section and input/output tapers.  (c) Calculated normalized field distributions ($E_x$ component) of the two fundamental supermodes (SM, symmetric, and AM, anti-symmetric) supported by the dual-core section at 1053-nm wavelength. The color code indicates the sign and strength of the electric field normalized to the maximum. The field distribution of the single-core mode TE$_\textrm{00}$ is shown as well. (d) Calculated group velocity dispersion (GVD) profiles. The SM (blue curve) possesses all-normal dispersion, while the TE$_\textrm{00}$ (black dashed curve) and the AM (orange curve) possess anomalous dispersion. With regard to single-core dispersion, the zero-dispersion wavelength (ZDW) of the anti-symmetric supermode is blue-shifted by approximately 50~nm.}
\label{fig:waveguides}
\end{figure}

\begin{figure}[t!]
\centering
\includegraphics[width=0.55\linewidth]{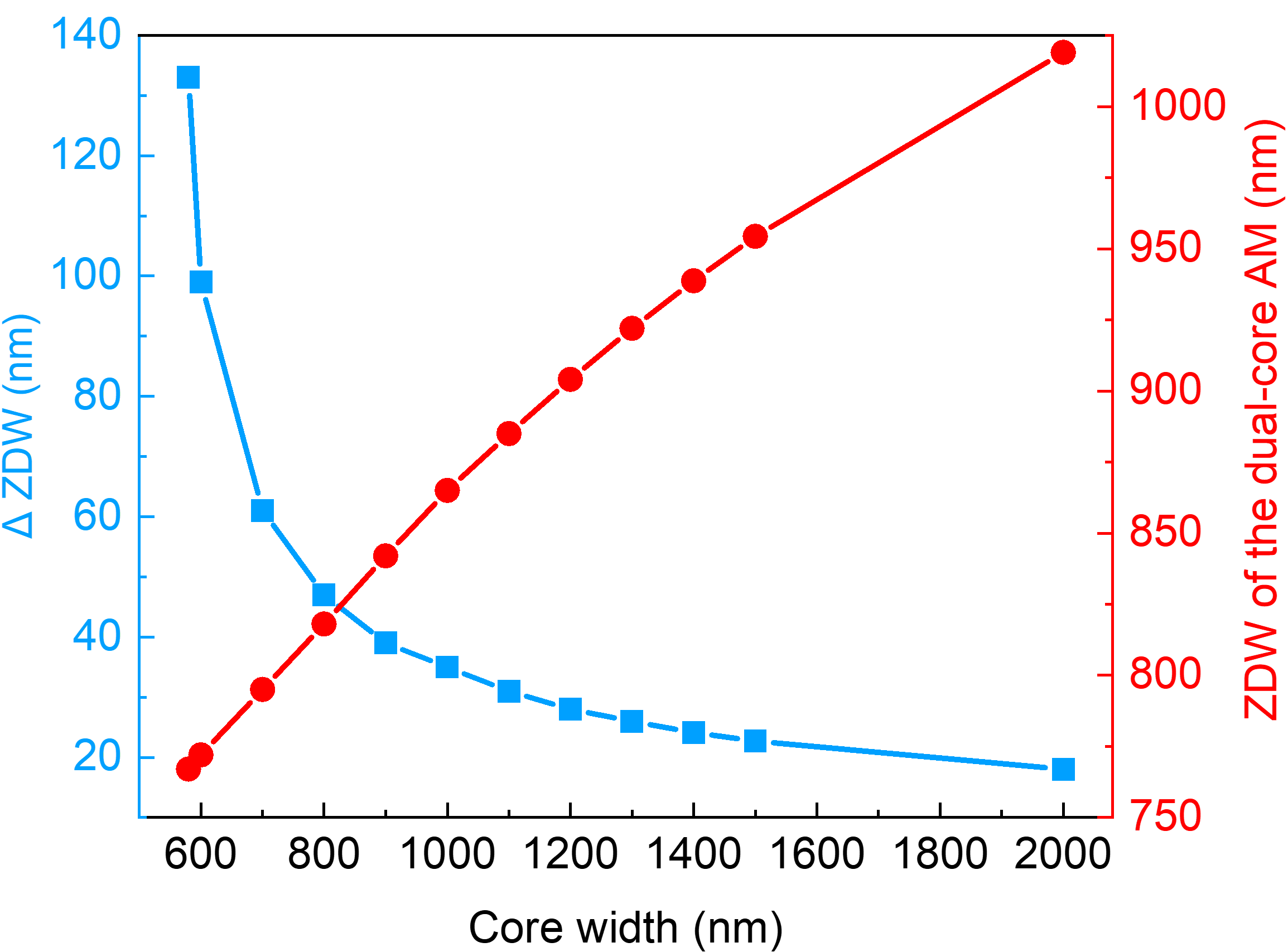}
\caption{$\Delta\textrm{ZDW}$ (blue curve) between the shortest ZDW of the dual-core waveguide for the AM and that of the corresponding single-core for the TE$_\textrm{00}$ mode, and absolute value of the ZDW (red curve) of the dual-core (AM) as a function of core width. The core height is fixed at 800~nm, and the gap between two cores for the dual-core structure is set to 300~nm. Core widths smaller than 580~nm are not included because the TE$_\textrm{00}$ mode no longer exhibits anomalous dispersion and does no longer have a ZDW.}
\label{fig:ZDW}
\end{figure}

Figure~\ref{fig:waveguides} (a) shows a schematic top view of the waveguide structure investigated in the experiments, fabricated by a Si$_{3}$N$_{4}$ foundry (LIGENTEC). For quantifying the frequency up-shift of dispersive waves accomplished via a dual-core waveguide cross section, and as reference, we also generate supercontinuum in a single-core waveguide (Fig.~\ref{waveguides} (b)) with the same core dimensions and fabricated adjacent on the same chip. For efficient coupling of the light in and out of the chip, both the dual- and single-core waveguides are equipped with 380-$\mu$m long inverse tapers designed for a mode-field diameter (MFD) of 2~$\mu$m at the nominal pump wavelength of 1053~nm. Due to the enlarged mode area, the characteristic nonlinear length in the tapers is in the order of a millimeter at the intensities used, which indicates negligible SC generation in the tapers. To allow for separately exciting single prongs, and to enable output measurements at single prongs, the prongs of the dual-core structure are well-separated with a transverse distance of 5~$\mu$m. For guiding the pump light into the 9-mm long dual-core section (colored in blue), short S-bends (60~$\mu$m) are designed for a bending loss orders of magnitude lower than the specified propagation loss of 0.2~dB/cm, at the nominal pump wavelength of 1053~nm.  We recall that single-prong excitation provides simultaneous excitation of the symmetric and anti-symmetric supermodes with equal amplitudes in the dual-core part of the structure~\cite{Burns1988,Ramadan1998ADC,Mrejen2015ADC}.

The dual-core waveguide section is designed to comprise two equal cores, each 0.80~$\mu$m by 0.80~$\mu$m in size, with a small, sub-wavelength horizontal gap of 300~nm between them which provides strong, wavelength dependent, coupling of the fields in the individual cores. For direct comparison, the core dimensions of the reference waveguide are the same. Mode analysis using a finite element 2D mode solver (Ansys Lumerical FDE) with material dispersion data~\cite{LIGENTEC} reveals that the dual-core cross section supports multiple supermodes, including the two lowest-order modes of interest, one with a symmetric and the other with an anti-symmetric field distribution, named SM and AM, respectively. In the following, these two modes will be referred to as the two fundamental modes. The transverse field distributions ($E_x$ component) of these two supermodes, calculated at a wavelength of 1053~nm, and the corresponding group velocity dispersion (GVD) profiles are shown in Figs~\ref{fig:waveguides} (c) and (d), respectively. For comparative analysis, also the shape and dispersion of the fundamental mode of the single-core waveguide is shown (TE$_\textrm{00}$, black dashed line). As can be seen in Fig.~\ref{fig:waveguides} (d), the dual-core AM provides a blue-shifted, shorter ZDW. To verify the generality of this claim, we fixed the waveguide core height at 800~nm and plotted the difference $\Delta\textrm{ZDW}$ (blue curve) between the shortest ZDW of the dual-core waveguide for the AM and that of the corresponding single-core for the TE$_\textrm{00}$ mode in Fig.~\ref{fig:ZDW}. In this figure the core width is varied and the gap  between the two cores is fixed at 300~nm. Figure~\ref{fig:ZDW} also shows the absolute value of the ZDW of the dual-core waveguide (red curve). From the figure it follows that reducing the core width results in both a larger $\Delta\textrm{ZDW}$ and a lower absolute value for the ZDW. Note, for core widths smaller than 580~nm, the single-core (TE$_\textrm{00}$) does not exhibit anomalous dispersion anymore and thus has no ZDW in the region of interest, while the dual-core waveguide still has one. Furthermore, the trend shown in Fig.~\ref{fig:ZDW} is also valid for other types of waveguides that are different in dimensions or material combinations, as exemplified in the conclusion.

\subsection{Experimental set-up \label{setup}}
\begin{figure}[b!]
\centering
\includegraphics[width=0.85\linewidth]{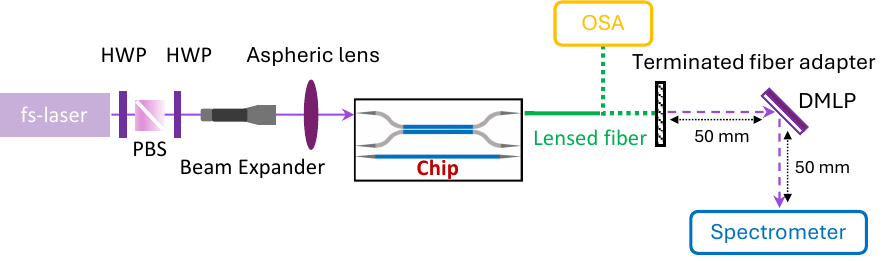}
\caption{Schematic view of the experimental set-up. Pump light is produced by a laser that operates at a center wavelength of 1053~nm with a 3-dB bandwidth of 21~nm, a pulse duration of 100~fs, and a repetition rate of 80~MHz. The laser is set at full power, and the pulse power is varied via the combination of two half-wave plates (HWP) and a polarizing beam splitter (PBS) in between. A beam expander and an aspheric lens are used to couple the laser beam into the input inverse tapers of the waveguides with a coupling loss of about 1~dB. The input aspheric lens is mounted on a piezo-controlled precision translation stage (Thorlabs MAX311D/M interfaced with a Thorlabs BPC203 piezo controller) for reproducible alignment and allows for changing the injection either between one of the two prongs of a dual-core waveguide or the inverse taper of an adjacent single-core waveguide. Lensed fibers (MFD of 2~$\mu$m) are used to collect output light from the waveguides for spectral analysis using an optical spectrum analyzer (OSA) or a spectrometer. When using the spectrometer to record the spectra shorter than 500-nm wavelength, the output of the fiber is sent over a free-space distance of 100~mm to the spectrometer, with a long-pass dichroic mirror (DMLP) placed at a free-space distance of 50~mm behind the fiber output. The purple line represents light guided in free space while the green line represents light guided in fibers.} 
\label{fig:setup}
\end{figure}

The experimental set-up used to investigate the shift in wavelength of dispersive waves generated by nonlinear optical (NLO) processes is schematically shown in Fig.~\ref{fig:setup}. Ultrashort pump pulses generated by a laser (Toptica Femto Ultra 1050), operated at full power with a center wavelength of 1053~nm (284.7~THz) and 100-fs pulse duration, travel through a built-in isolator. A half-wave plate (HWP) and a polarizing beam splitter (PBS) followed by another HWP is used to adjust the pulse energy sent towards the waveguides. Next, the laser beam passes through a 2.5-times achromatic beam expander (Thorlabs ZBE1C AR Coated: 1050-1650~nm) and is focused with an aspheric lens (Thorlabs C330 TMD-C, f=~3.1~mm at 1053-nm wavelength) into either a single input prong of the dual-core waveguide, or into the single-core reference waveguide. We calculated that chirp due to the group delay dispersion of the input coupling optics can be neglected (GDD less than $10^{-3}$ ps$^2$). It is important to note that our primary interest lies in exciting the anti-symmetic supermode (AM) of the dual-core structure. However, for convenience and easy input coupling, injection of the pump pulses into a single input prong was chosen, leading to a simultaneous and equal excitation of the symmetric and anti-symmetric supermodes (SM and AM) at the start of the dual-core waveguide section. We note that this requires doubling the pump power to obtain the same excitation for the AM as for the $\textrm{TE}_{00}$ mode. Channeling all power into a single supermode in a next step may rely, e.g., on a tunable Mach-Zehnder interferometer (MZI) input circuit as described in ~\cite{Xia23}. We note that co-excitation of the SM is not a concern in the present work as the two modes (SM and AM) quickly decouple upon propagation (see section~\ref{Modeling}). For spectral analysis, output light is collected using lensed fibers (MFD 2~$\mu$m, OZ Optics TSMJ-3A-1550-9/125-0.25-7-2-14-2-AR-SP) placed behind one of the prongs. This corresponds to observing a superposition of light from both supermodes as explained in~\cite{Xia25}. Spectral measurements from the near-IR to the mid-visible range (500 to 1750~nm) are done with an optical spectrum analyzer (OSA, ANDO AQ6315A,  50-pm resolution). For mid-visible to UV recordings (350 to 500~nm) a spectrometer is used (OceanInsight FLMT08730 with a 10-$\mu$m input slit, corresponding to a 1.0-nm resolution~\cite{OceanOptics2025}). For the latter, the fiber output is directed to the spectrometer over a free-space distance of approximately 100~mm. This distance allows to insert a dichroic mirror (Thorlabs DMLP567) as long-pass filter that transmits wavelengths longer than about 570~nm to a beam dump, while reflecting the shorter-wavelength components into the spectrometer. The filtering was used to avoid saturation of the detector with the dominant infrared radiation around the pump wavelength. To optimize the input coupling and output detection, the transmission at the pump wavelength is maximized using low pump power to ensure that nonlinear conversion is negligible. For direct observation of visible-light generation, an RGB camera and a microscope (Olympus SZ-CTV) are used to take top-view pictures of the chip.

\subsection{Numerical modeling \label{Modeling}}
For a later comparison  of the experimental results with theoretically expected spectra, thereby guiding the identification of underlying processes that generate visible light,  two numerical models are employed, in order to include all of the possibly involved spatial and spectral properties: the first is capable to model spatially multimode dynamics, based on the multimode generalized nonlinear Schrödinger equation (MM-GNLSE)~\cite{Poletti2008DescriptionFibers,NiklasandMaximilian2021Doi10.1002/lpor.202100125}, while the second model is limited to a single mode but includes also third harmonic generation (THG) ~\cite{voumard2023simulating}.

\begin{figure}[b!]
\centering
\includegraphics[width=0.5\linewidth]{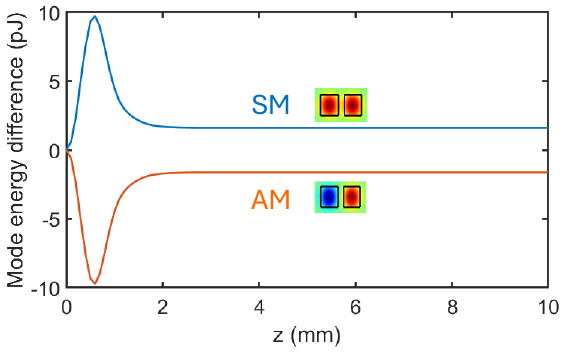}
\caption{Calculated change of mode energy in the SM and the AM versus propagation length for simultaneous excitation with equal pump energies of 375~pJ. }
\label{fig:ChangeModeEnergy}
\end{figure}

More specifically, the MM-GNLSE allows to simulate SC generation involving multi-mode excitation and interaction, such as might be required to describe SC generation in the dual-core waveguide. The MM-GNLSE model used in the work includes dispersion up to 15$^\text{th}$-order, self-steepening, and nonlinear coupling of the modes. We disabled Raman scattering because this effect is a very weak process in SC generation in Si${_{3}}$N${_{4}}$ waveguides~\cite{klenner2016gigahertz}. The SM and AM in the dual-core waveguide are expected to decouple after a very short propagation distance $L_p = \bar{v}_g^2\tau_p/\Delta v_g$,  where $\bar{v}_g$ is the mean group velocity at the pump wavelength, $\tau_p$ the pulse duration and $\Delta v_g$ the difference in group velocity. Using $\bar{v}_g = 1.4 \times10^{8}$~m/s, $\tau_p = 100$ fs and $\Delta v_g = 1.7 \times 10^{6}$~m/s yields a propagation distance of $L_p = 1.2$~mm after which the temporal overlap between the pump pulse in the two supermodes is expected to be gone. As a more quantitative test, we numerically simulated the change of mode energy vs propagation distance under simultaneous excitation of both modes, each with a pulse energy of 375~pJ corresponding to the maximum used in our experiments.  Figure~\ref{fig:ChangeModeEnergy} shows that the energy exchange is rather weak, settling at below 1~$\%$ of the energy, and ceases before a propagation distance of 1.8~mm. Therefore, also a single-mode model should be able to describe the dynamics of the system. 

The second, single-mode, model~\cite{voumard2023simulating} allows to simulate  mixed and cascaded nonlinearities which typically requires a much broader spectral simulation window. The model uses the smallest possible window that allows the out-of window generated light to fold back into the simulation window, while remaining spectrally separated from the physical spectrum already present. This makes the model computationally efficient. In the remainder, this model is referred to as single-mode cascaded generalized nonlinear Schrödinger equation (SMC-GNLSE). Dispersion up to the 15$^\text{th}$ order is included, while the second-order susceptibility is set to zero, owing to the amorphous structure of the waveguide materials.

The models use chirp-free input pulses as in the experiment, while a sech$^2$ temporal shape is assumed for simplicity. The pulse energies applied in the simulation match the estimated experimentally in-coupled energies. To allow a direct comparison, we chose to model SC generation using the same pulse energy either in the AM, the SM or the TE$_\textrm{00}$ modes. The simulation is performed over a propagation length of 10~mm. 

\begin{figure}[t!]
\centering
\includegraphics[width=0.5\linewidth]{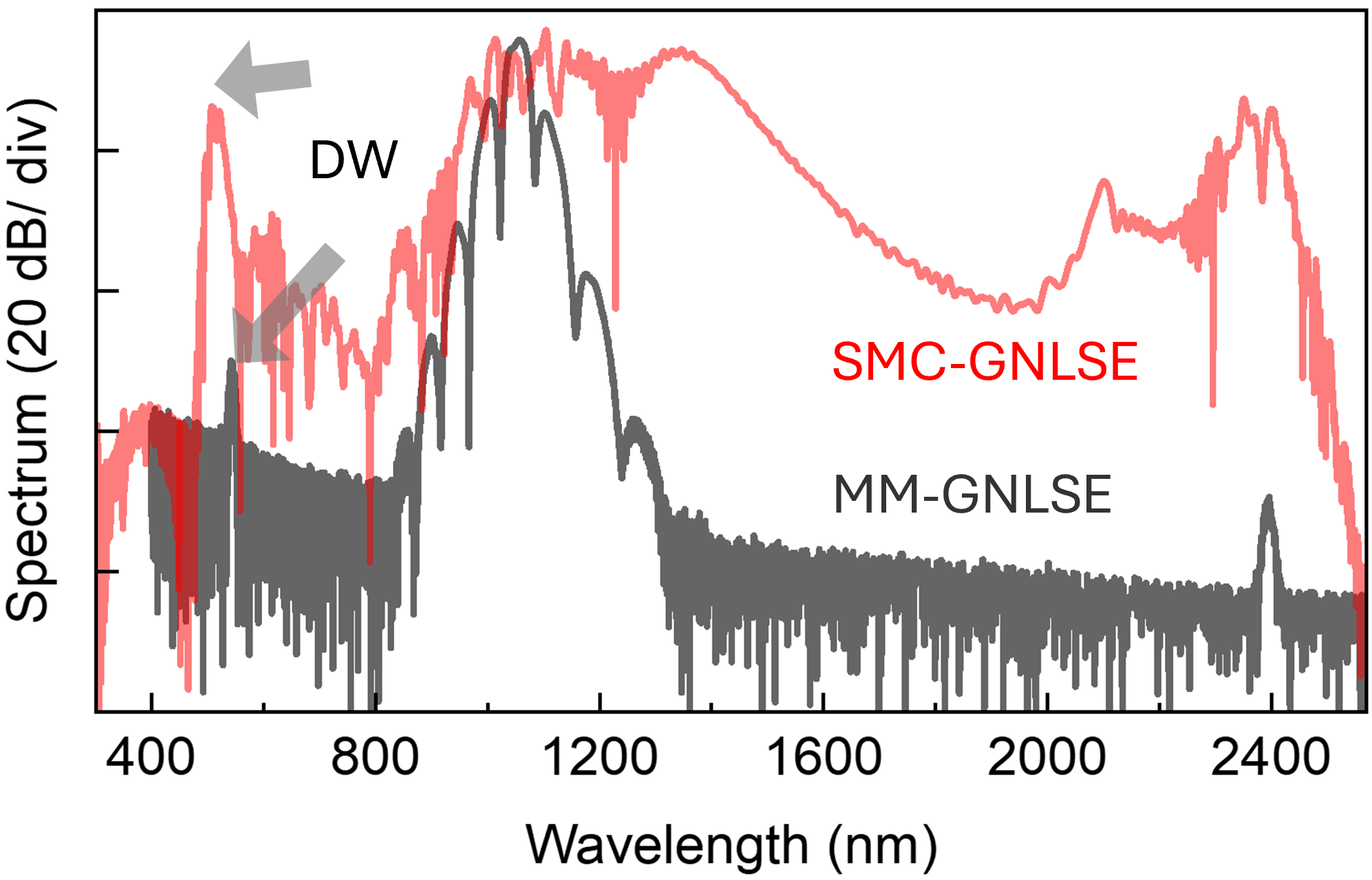}
\caption{Comparison of the calculated spectra by MM-GNLSE with solely AM excitation (black curve) and the SMC-GNLSE (red curve) for an input pulse energy of 187.5~pJ. All physical input parameters are the same. DW: dispersive waves.}
\label{fig:ModelComparison}
\end{figure}

For comparison between the two models we use the same values for the physical input parameters, which include the mode dispersion ($n_\textrm{eff}$), and use the MM-GNLSE with solely AM excitation. Figure~\ref{fig:ModelComparison} shows the calculated spectra by both models for an input pulse energy of 187.5~pJ.  It can be seen that, although some overall spectral features agree well, the location of the short-wavelength DW differs by about 8\%. Furthermore, the SMC-GNLSE model shows more structure. We find that the MM-GNLSE model shows better agreement with our measurements at wavelengths down to about 500~nm, while the SMC-GNLSE model predicts light generation at shorter wavelengths, which is also observed in our measurements. In order to capture also any transient and small inter-modal interaction, and because both supermodes are excited in the experiment, we use the MM-GNLSE to compare with our measured data. As this model does not include the physics of third harmonic generation, we use the SMC-GNLSE model to test for the presence of this process.

\section{Results and discussions}
\subsection{Wavelength shift in dispersive waves}
\label{NIR broadening}

\begin{figure}[b!]
\centering
\includegraphics[width=\linewidth]{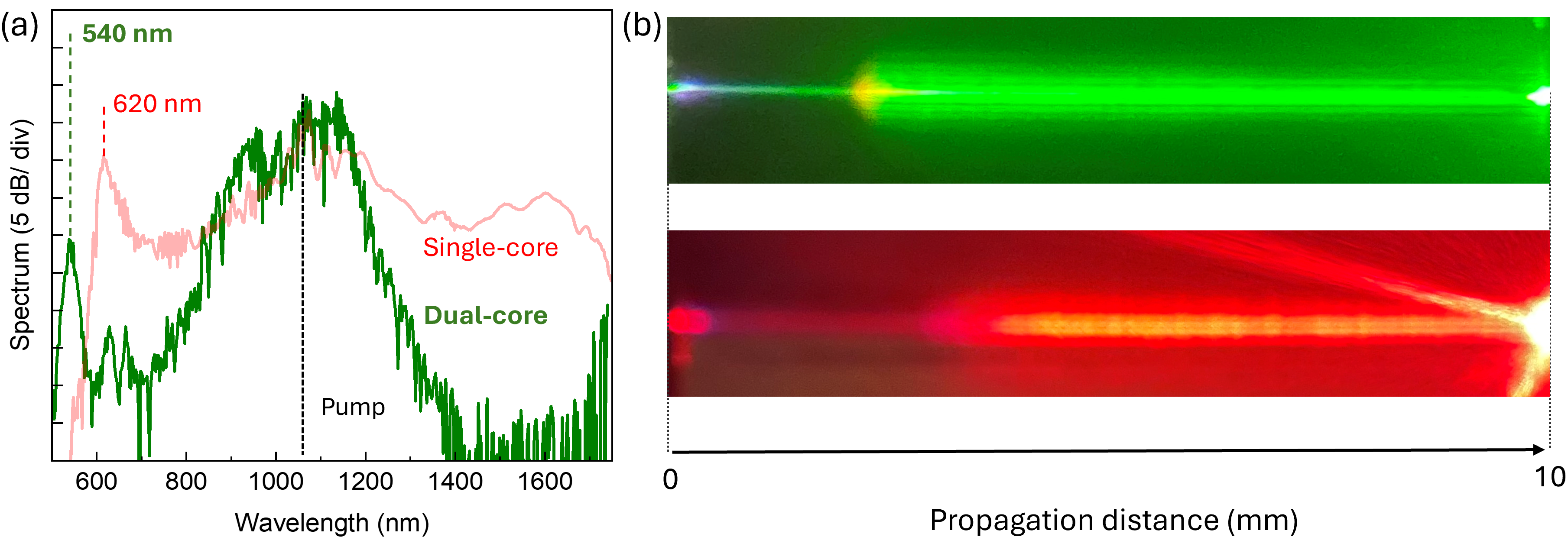}
\caption{(a) Measured supercontinuum (SC) spectra for the dual-core (green) and the single-core (red) waveguides. For the dual-core waveguide, SM and AM are simultaneously excited, while for the single-core waveguide only the TE$_\textrm{00}$ mode is excited. The applied pump energy is the same (375~pJ) in each of the SM, AM and TE$_\textrm{00}$ modes. (b) Top-view photographs show that green light is generated with the dual-core waveguide whereas the single-core waveguide generates red light.}
\label{fig:Spectra_Measured}
\end{figure}

Figure~\ref{fig:Spectra_Measured} (a) shows the measured SC output spectra generated with simultaneous excitation of SM and AM in the strongly coupled, dual-core waveguide (green trace), and with TE$_\textrm{00}$ excitation in the single-core waveguide (red trace). The same pulse energy of 375~pJ is injected into each spatial mode (i.e., a total of 750~pJ for the dual-core waveguide). Such pulse energies were selected because we observed that at this level the spectral evolution became well-completed in all cases, i.e., the spectral width of the output did not increase with higher pulse energy neither in the dual-core nor single-core waveguides.  It is important to note that for the dual-core waveguide output, we observed no noticeable difference in the measured spectra from the two output prongs, except around the pump wavelength where a rapid spectral oscillation is found with the same spectral envelope as the two contributing components (SM and AM). This indicates that further away from the pump wavelength, a significant amount of light is present only in one of the supermodes because if light were present in both the SM and AM supermodes, their interference would result in a rapid spectral oscillation in the output signal between the two output prongs, as observed in~\cite{Xia25}. A striking difference between the outputs from the dual-core and single-core waveguides can be seen from the top-view images, as shown in Fig.~\ref{fig:Spectra_Measured} (b). Specifically, the dual-core waveguide emits green light, whereas the single-core waveguide emits in the red spectral range. The spectra show that the output SC spectrum (green curve) of the dual-core waveguide has, except for the dispersive wave, a smaller bandwidth than the single-core SC (red curve). However, the dual-core SC has a shorter wavelength peak located at 540 nm that is blue-shifted by almost 80 nm compared to that of the single-core SC. 

\begin{figure}[t!]
\centering
\includegraphics[width=\linewidth]{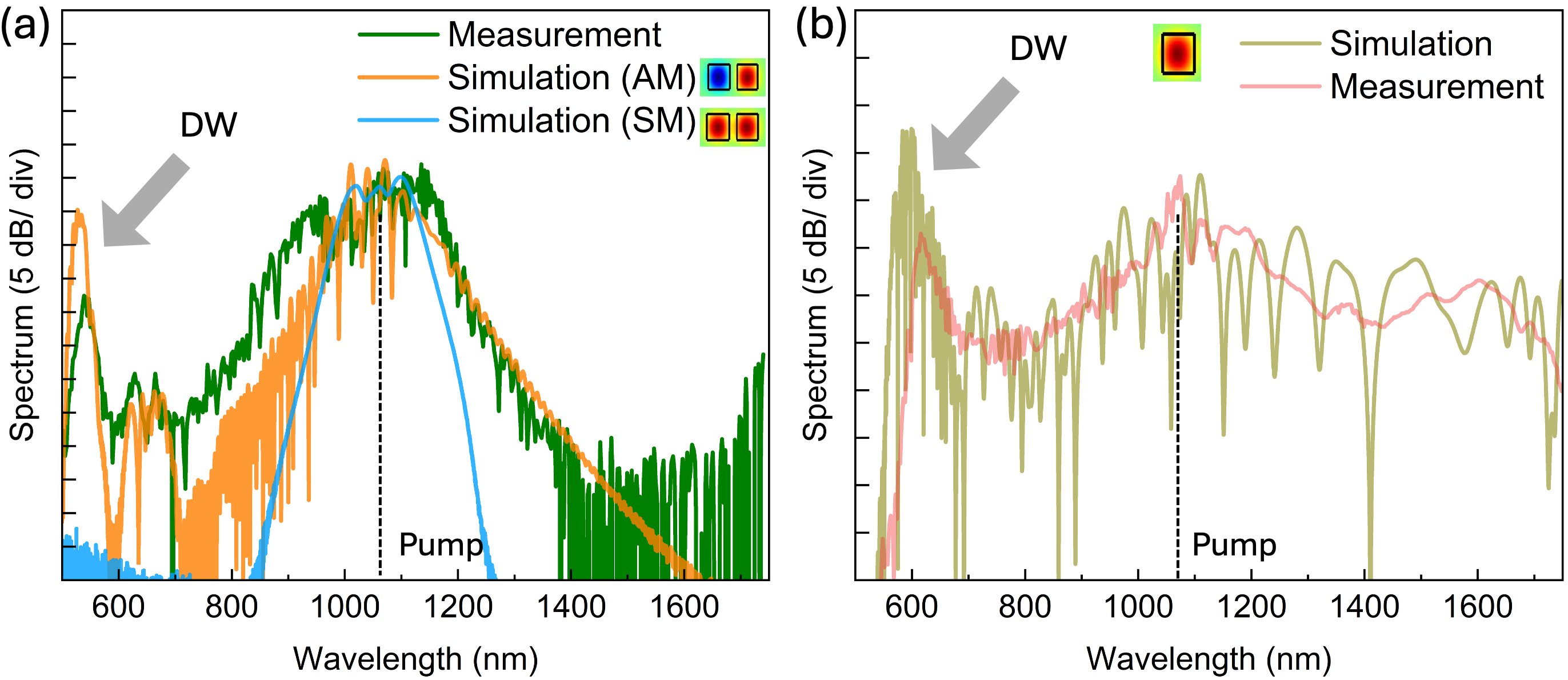}
\caption{(a) Comparison of the measured SC spectrum with simulated spectrum for simultaneous SM and AM excitation. Except for an overall factor related to the detection efficiency, no fit parameters were used. Narrowband green emission around 540~nm (green trace) matches well with the prediction of dispersive wave (DW) generation in the anti-symmetric supermode, AM. In contrast, SC generation in the symmetric supermode, SM, comprises only infrared self-phase modulation. (b) The measured and calculated SC spectra also agree well for SC generation in the single-core waveguide (single TE$_\textrm{00}$ mode), however, the visible dispersive wave is found at a longer wavelength near 620~nm. The MM-GNLSE model is used for the simulations. The in-coupled pump pulse energy is identical in both the experimental measurements and numerical modeling, with a value of approximately 375~pJ for each of the SM, AM, and TE$_\textrm{00}$ modes. The simulation results for both waveguides show that the location of the dispersive wave is predicted with an accuracy better than 2\%. }
\label{fig:Spectra_Simulated}
\end{figure}

For a more comprehensive comparison, we calculated the output spectra for the dual-core and single-core waveguides using the MM-GNLSE model, as shown in Figs.~\ref{fig:Spectra_Simulated} (a) and (b), respectively. As in the experiment, both supermodes are initialized with equal pulse energies. As expected, the SC spectra generated in the SM and AM have distinctly different shapes due to an opposite sign of dispersion.

Comparison between the measured SC spectrum of the dual-core waveguide with the simulated spectra shows that for wavelengths far from the pump (beyond approximately an interval spanning 900 to 1200~nm), SC is dominantly generated in the AM, i.e., contributions from SM supercontinuum are negligible. This suggests that the spectral components of the measured SC spectrum at these wavelengths originate solely from the AM excitation. This aligns with the earlier conclusion that only a single supermode contains light far from the pump wavelength, drawn from the experimental observation of no spectral difference between the two individual output prongs.

\begin{figure}[t!]
\centering
\includegraphics[width=\linewidth]{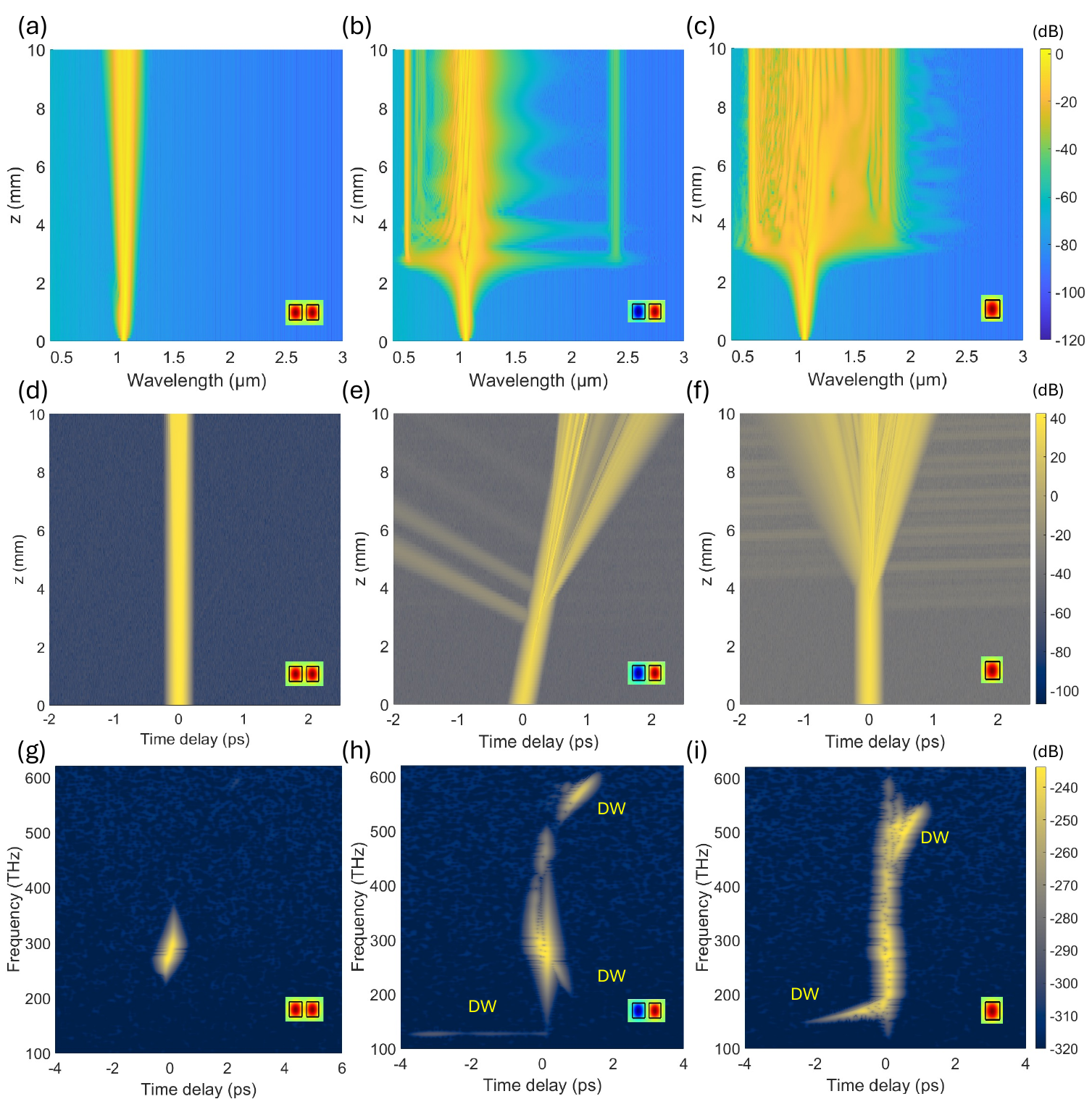}
\caption{Simulated evolution of SC spectra in (a) the SM and (b) the AM of the dual-core strongly coupled waveguide, and in (c) the TE$_\textrm{00}$ of the single-core waveguide. Simulated temporal evolution of SC spectra in (d) the SM and (e) the AM of the dual-core waveguide. The temporal window is chosen to move with the group velocity of the the pump pulse in the SM. (f) Temporal evolution of SC spectrum of the TE$_\textrm{00}$ of the single-core waveguide. Simulated spectrogram of SC generation in (g) the SM and (h) the AM of the dual-core waveguide and (i) the TE$_\textrm{00}$ of the single-core waveguide at the output of the waveguides. The MM-GNLSE model is used for the simulations. Each mode has the same pulse energy, equal to 375~pJ. DW: dispersive waves.}
\label{fig:1050Modelling}
\end{figure}

Comparing further, the narrow spectral peak centered around 540~nm, which forms the short-wavelength edge in good agreement with the simulation, is identified as a dispersive wave (DW). We also verified this via the standard phase-matching equation for dispersive wave generation~\cite{Agrawal_6th_Nonlinear_fiber_optics}, as well as via the simulated spectral evolution and spectrogram (see Fig.~\ref{fig:1050Modelling} and discussion below). Also, the visible red spectral peak found as the short-wavelength edge in the measured single-core SC spectrum is well predicted (Fig.~\ref{fig:Spectra_Simulated} (b)).


The difference in the central part of the supercontinuum for the two waveguides, as observed in both experiments and modeling, is attributed to the difference in soliton order generated when using the same pump energy per mode. Namely, the soliton order $N$ is given by $N^2 ={\gamma P_0 T_0^2}/{|\beta_2|} $, where $\gamma = {n_2 \omega_0}/{c A_{\text{eff}}}$ is the nonlinear coefficient, $n_2$ is the Kerr index, $\omega_0$ is the angular frequency of the pump pulse, $P_0$ is the peak power of the pulse, $T_0$ is the pulse duration, $c$ is the speed of light in vacuum, $A_\text{eff}$ is the effective mode area, and $\beta_2$ is the GVD parameter at the pump wavelength. As the $\beta_2$ and the $A_\text{eff}$ are both about twice as large for the dual-core waveguide, the soliton order is lower by a factor of 2 in the dual-core waveguide.

For gaining deeper insight into the underlying physical mechanisms we present and compare the simulated nonlinear dynamics of the three experimentally investigated modes, i.e., the AM and SM of the dual-core waveguide and the single-core TE$_\textrm{00}$ mode. The spectral evolution along the propagation coordinate (z) is displayed in Figs.~\ref{fig:1050Modelling} (a) to (c), respectively, the corresponding evolution of the temporal shape is shown in Figs.~\ref{fig:1050Modelling} (d) to (f) and the spectrogram at the output of the waveguide in Figs.~\ref{fig:1050Modelling} (g) to (i), for completeness. 
For the SC spectrum in the SM, given the all-normal dispersion, it can be seen that SPM governs the dynamics throughout the entire 10-mm propagation length. This is apparent as smooth, but limited, spectral broadening, while dispersion slowly increases the pulse duration.  
In contrast, for the SC spectrum in the AM, SPM is the main contributor only at the beginning of propagation, and at around 2~mm a higher-order soliton emerges. With further propagation at a distance of about 2.5~mm, soliton fission appears that splits the compressed pulse into several fundamental solitons traveling with different velocities. Ruled by phase and group velocity matching, two distinct DWs are generated at the shortest and longest wavelengths, which lie in the green spectral range around 540~nm and in the mid-IR at 2.4~$\mu$m, respectively. The SC spectrum generated in the TE$_\textrm{00}$ mode of the single-core reference waveguide is also driven by soliton fission, as in the AM. Here, however, the dynamics is slightly slower and fission takes place only after a propagation distance of 3~mm, generating DWs at 620~nm and 1.8~$\mu$m. The propagation distances towards soliton-fission found in simulations correspond well to what is observed in the experiments, i.e., where visible scattered light shows up in top-view camera recordings (see in Fig.~\ref{fig:Spectra_Measured} (b), approximately at 2.5~mm and 3~mm, respectively). Regarding the generation of dispersive waves, the DWs in the dual-core AM exhibit rapid walk-off from the pump pulse tail, unlike what occurs in the single-core (TE$_\textrm{00}$), as seen in Figs.~\ref{fig:1050Modelling} (e) and (f), respectively. Interesting to note in the spectrogram in Fig.~\ref{fig:1050Modelling} (h) is that the trailing short-wavelength DW of the dual-core waveguide may be separated from the remainder of the SC if using a spectral filter. Furthermore, this DW is chirped and could thus be temporally compressed. On the other hand, the corresponding DW generated in the single-core waveguide (see Fig.~\ref{fig:1050Modelling} (i)) cannot easily be separated from the remaining SC. 

In the discussion above, we have highlighted the advantage of generating SC in the dual-core waveguide over a single-core waveguide, which is enabling a spectral shift of dispersive waves toward shorter wavelengths. On the other hand, the single-prong excitation used in our experiments reduces the efficiency of DW generation by a factor of two, as the SM does not contribute to the visible light generation. However, this disadvantage can be removed in the future by modifying the pump injection scheme. Specifically, when identical pump pulses are injected in both prongs simultaneously, though with opposite optical phase, all pump radiation would be diverted into the anti-symmetric supermode, AM. 

One may consider that the higher-order $\text{TE}_{01}$ transverse mode in a single-core waveguide might also be effective in shifting the ZDW to higher frequency, because its field profile is quite similar to that of the dual-core AM if the single-core waveguide width is taken the same as the total width of the dual-core waveguide. Indeed, this is true, but only for very small gap to core ratios ($\ll 1$). For such  ratios, we expect the GVD of the two modes to be quite similar and have nearly identical ZDW. This has been confirmed by calculating the GVD for the two cases. On the other hand, Fig.~\ref{fig:ZDW} shows that larger ratios are desirable for generating shorter-wavelength DWs. For ratios $\geq 0.5$ the GVD for the $\text{TE}_{01}$ mode starts to deviate, leading to a larger GVD at the pump wavelength and a larger ZDW compared to the dual-core waveguide. Thus in combination with the simpler on-chip excitation circuit, the dual-core waveguide geometry is preferred.

\subsection{UV light generation \label{UV_intro}}
\begin{figure}[b!]
\centering
\includegraphics[width=0.45\linewidth]{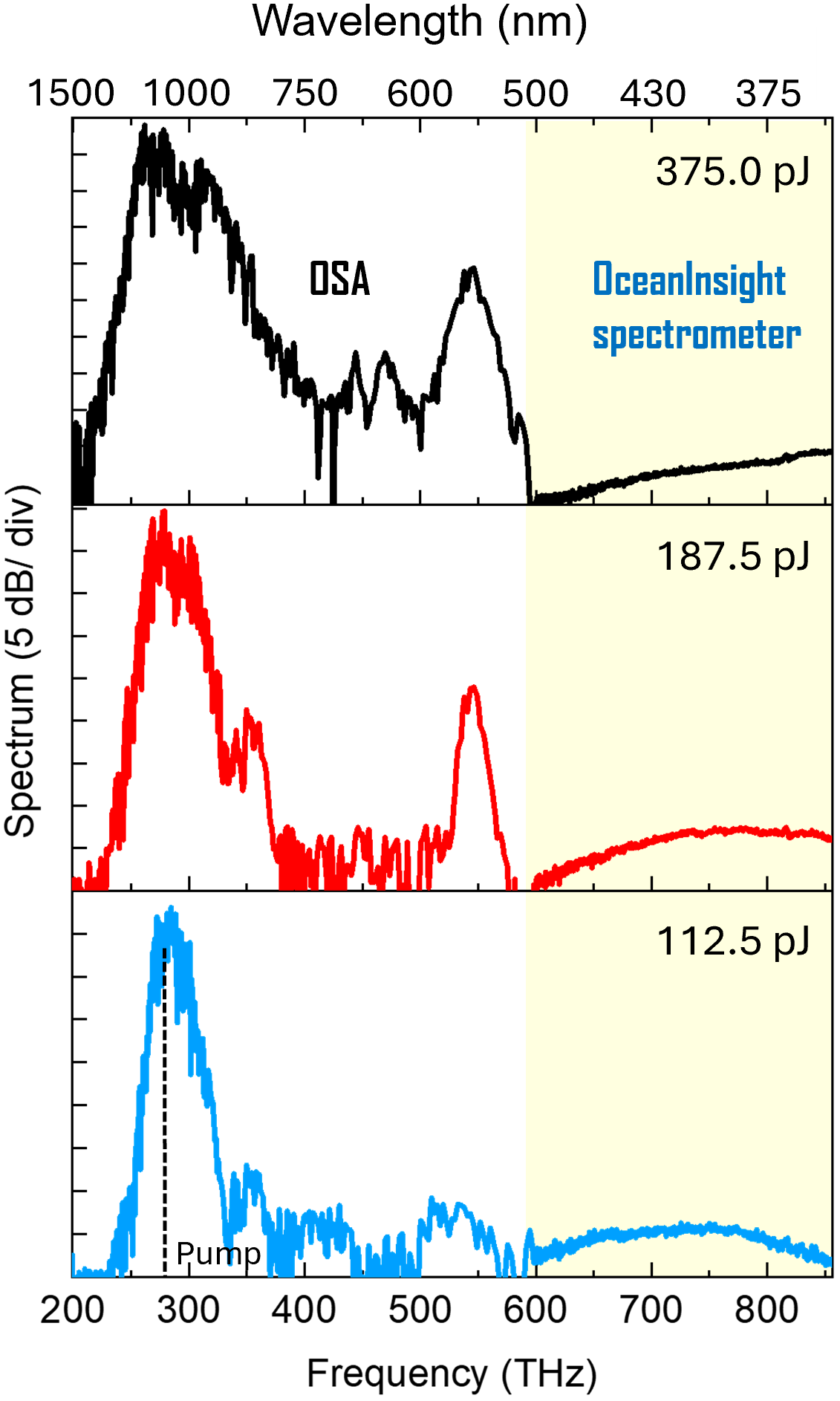}
\caption{Complete spectral profile as a function of optical frequency for different in-coupled energies per mode. The corresponding optical wavelengths are indicated on the top axis for reference. The spectral region below approximately 600~THz was measured using the OSA, while the region above 600~THz was obtained with the OceanInsight spectrometer (yellow area). Note: spectral data has not been corrected for frequency-dependent transmission losses.}
\label{fig:UV}
\end{figure}

\begin{figure}[b!]
\centering
\includegraphics[width=0.55\linewidth]{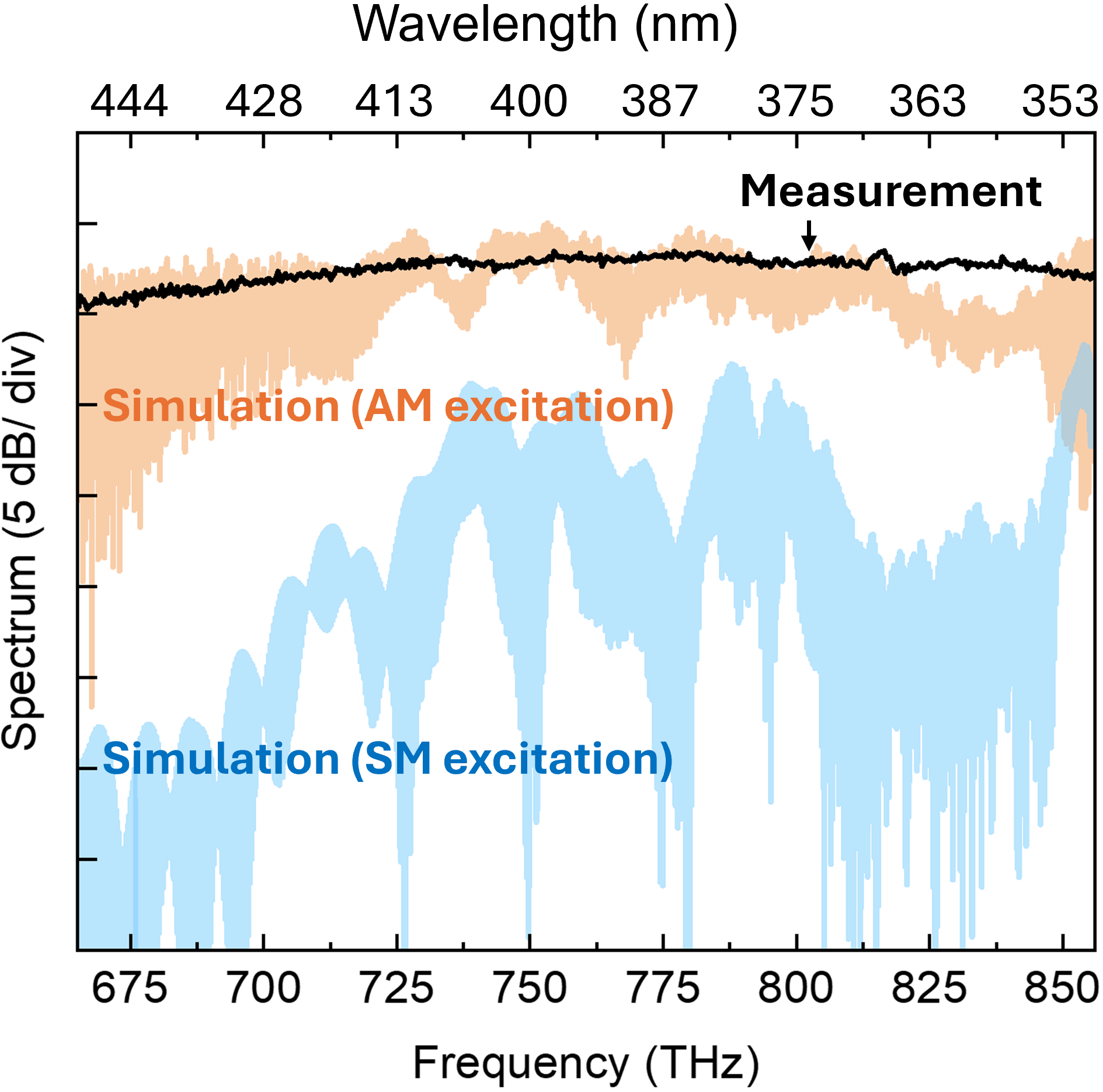}
\caption{Comparison of the measured blue-UV spectrum (black curve) with the simulated blue-UV spectra under solely AM excitation (orange curve) and solely SM excitation (blue curve) as a function of optical frequency. Corresponding wavelengths are labeled on the top axis. The SMC-GNLSE model is used for the simulations. Each mode has the same pulse energy, equal to 187.5~pJ.}
\label{fig:UV_simulation}
\end{figure}

Turning to the measurements using also the second, shorter-wavelength spectrometer, we observe that the dual-core waveguide not only facilitates frequency up-shifting of the dispersive wave into the green spectral range, but also that light is generated in the blue and UV ranges for in-coupled pulse energies in the waveguide at and above approximately 225~pJ. Such generation was not observed with the single-core waveguide at any of the pump pulse energies used in the experiments. To investigate the mechanism underlying the UV generation, a comprehensive view is essential as nonlinear processes involve interactions among a broad range of spectral components.  We therefore constructed a complete spectral profile as a function of light frequency, shown in Fig.~\ref{fig:UV}, by scaling and merging the OSA-measured spectra and those acquired by the spectrometer. The top axis provides the corresponding wavelength values to facilitate interpretation across both domains. 

We note that merging the profiles only provides an approximate overview, as frequency-dependent loss in the detection path, e.g., caused  by the finite transparency range of the waveguide or in the collection by the lensed fiber, was not taken into account. Nevertheless, we verified that the measured spectra are representative for the spectral content of the light emitted by the output prongs. This was done by repeated re-doing and checking the alignment,  intentionally perturbing the fiber, or by collecting the light from the other output prong, all of which yielded comparable spectra.

Figure.~\ref{fig:UV} shows a systematic change of the spectra versus increasing pump pulse energy in that it increases the output power in the blue-UV band, while also shifting the center of the band to higher frequencies. While the center frequency lies at about 720~THz (light wavelength around 415~nm) with 112.5~pJ pulse energy per mode, it shifts to above 856~THz (below 350~nm) at 375~pJ per mode, exceeding the detection range of the spectrometer. In parallel, the OSA-spectrum shows that a distinct dispersive wave at 540~nm (light frequency of 555~THz) is generated only at pump energy levels higher than about 112.5~pJ per mode. 

Because the observed UV light is approximately at the third harmonic of the pump, we suspect that this light is generated by third harmonic generation. As this process is not included in the MM-GNLSE model, we use the SMC-GNLSE model~\cite{voumard2023simulating} for further analysis. We recall that this model simulates light propagation and evolution in a single mode only. Nevertheless, it can still be employed here and remains valid as discussed in section~\ref{Modeling}.

Figure~\ref{fig:UV_simulation} shows measured and simulated spectra for the frequency range that is given by the band edge of the filter and the shortest spectrometer wavelength. Simulated spectra are shown for both the dispersion of the AM (orange) and SM (blue) modes, each having a pulse energy of 187.5~pJ. This intermediate pulse energy was chosen because important spectral features remained within measurement range. The model predicts third harmonic generation of the broadened pump but much of the generated third harmonic falls outside the range of the spectrometer. What is shown in Fig.~\ref{fig:UV_simulation} is THG only from frequencies lower than the central pump frequency.  Comparing the two simulated spectra (orange vs blue curves), it can be seen that the AM excitation leads to approximately 10~dB more intense blue-UV radiation than the SM excitation, from which we conclude that the measured spectrum in blue-UV range is predominately attributed to the anti-symmetric supermode, AM.

The observation that the third harmonic intensity is typically 25~dB below that of the DW indicates a non-phase-matched THG process. This is confirmed also by calculating a phase mismatch of several hundred $\pi$. Another contributing factor may be strong material absorption in the UV.


\section{Conclusion and outlook}
To conclude, we demonstrate an effective way to realize on-chip shifting of visible-range dispersive waves toward shorter wavelengths, which is straightforward to implement. The approach is based on supermode dispersion engineering in multi-core waveguides. In the particular system investigated here, by straightforwardly placing a second waveguide core adjacent to the original single-core waveguide, we facilitated a dispersion profile with a blue-shifted ZDW, leading to dispersive wave generation at shorter wavelengths.

We have shown that blue-shifting the zero-dispersion wavelength via dual-core waveguides requires small core widths (Fig.~\ref{fig:ZDW}). The short-wavelength DW will also blue-shift, the actual amount being determined by the phase matching condition for dispersive wave generation under specific experimental settings. For example, for a core width of 800~nm, the ZDW is calculated to be at 818~nm and the DW is experimentally observed at 540~nm.

In order to illustrate the general applicability of this approach, we have also considered different pump wavelengths and other waveguide geometries. For example, using a longer pump wavelength of 1550~nm and the same dual-core waveguide geometry (0.8- by 0.8-$\mu$m cores with 300-nm gap), the short-wavelength DW is calculated to shift from 580~nm to 425~nm. 
Also, using uncladded Si$_{3}$N$_{4}$ waveguides as in ~\cite{Tagkoudi21}, the DW is calculated to shift from 550~nm to 515~nm when using two cores with a 150-nm gap.

Finally, the dispersion of the dual-core waveguide facilitates non-phase matched THG via the third-order susceptibility, which was not observed in the corresponding single-core waveguide. This contrast was also observed in a similar dual-core waveguide having a core width of 1~$\mu$m \cite{Xia25}. Optimizing the dispersion, for example, by using cores with different cross sections may enhance this process.

Potentially, the approach and the results in this work can provide an extra perspective for developing compact integrated visible-UV light sources.


\section*{Funding} This work is partially funded by the Dutch Research Council (NWO) SYNOPTIC OPTICS program, P17-24 project 1.

\section*{Acknowledgments} The authors would like to thank C. Toebes for providing the OceanInsight spectrometer.

\section*{Disclosures} The authors declare no conflicts of interest.

\bibliography{Citation}

\end{document}